\documentclass[aip,pop,amsmath,amssymb,reprint]{revtex4-2}

\usepackage{graphicx}
\usepackage{amsmath}
\usepackage{bm}
\usepackage{color}

\begin{document}

\title{Complex plasma with Janus particles as a model active-matter system}

\author{V. Nosenko}
\email{V.Nosenko@dlr.de}
\affiliation{Institute for Frontier Materials on Earth and in Space, German Aerospace Center (DLR), D-51170 Cologne, Germany}
\affiliation{Center for Astrophysics, Space Physics, and Engineering Research, Baylor University, Waco, Texas 76798-7310, USA}

\date{\today}
\begin{abstract}
Active matter classifies systems consisting of self-propelled units which convert the energy stored locally or extracted from their environment into directed motion. It has recently attracted considerable attention due to rich new physics it displays and potential applications in various fields including materials science. Active matter found in nature is inherently complex, so model systems are of interest where the main relevant features can be isolated and studied in laboratory experiments. An interesting instance of active matter is a suspension of active particles (e.g., the so-called Janus particles, where the two halves have different properties) in a gas discharge plasma. Such complex plasmas with active particles are excellent model systems which can enhance our understanding of natural active matter systems not easily amenable to experiment. In the present experimental study, micrometer-size plastic microspheres with thin gold coating on one side were suspended as a single layer in the plasma sheath of a radio-frequency discharge in argon and driven by a combination of laser-induced photophoretic force and asymmetric ion drag force. Enhanced particle activity in this highly driven, inertial active-matter system leads to collective particle dynamics characterized by extended self-similarity of the velocity field, intermittency, and the emergence of a direct energy cascade with non-universal scaling exponent.
\end{abstract}


\maketitle

\section{Introduction}

Active matter systems are ensembles of self-propelled units which convert the energy extracted from their environment or stored locally into directed motion, thereby maintaining the system in a nonequilibrium state \cite{Elgeti:2015,Bechinger:2016}. Such systems exhibit collective phenomena that have no direct equilibrium analogs and are of interest across a wide range of disciplines, from soft condensed matter physics to materials science and biology. Because naturally occurring active matter systems often involve many coupled processes, simplified experimental realizations are essential for isolating key physical mechanisms under controlled conditions.

A particular instance of active matter is a suspension of active particles in a gas-discharge plasma known as complex, or dusty plasma \cite{Ivlev_book}. The particles receive large electric charges due to collection of electrons and ions from plasma. The charge is usually negative because of the prevalence of electron current upon initial contact with plasma. Through their interaction and external confinement the particles self-organize into structures with liquid-like or solid-like ordering. A defining feature of complex plasmas is that the particle motion can be resolved at the level of individual constituents in real time, enabling direct access to microscopic dynamics and correlations. Owing to the typically low neutral gas damping, particle inertia plays an important role, distinguishing complex plasmas from overdamped model systems such as colloidal dispersions \cite{Ivlev_book}. This has enabled detailed studies of transport processes \cite{Nunomura:2006,Nosenko:04PRL_visc,Gavrikov:2005,Hartmann:2011,Nosenko:08PRL_therm}, phase transitions \cite{Thomas:1996,Nosenko:2009,Melzer:2013}, and collective excitations including waves and instabilities \cite{Nunomura:2002,Piel:2002,Zhdanov:2003,Avinash:2003,Couedel:2010}.

As was recently shown, the so-called  Janus particles---microspheres with deliberately introduced surface asymmetry \cite{Bechinger:2016,Walther:2013}---become self-propelled when suspended in radio-frequency argon plasmas \cite{Nosenko:2020PRR_JP, Arkar:2021, Nosenko:2022, Nosenko:25PRE}. In these systems, directed motion arises from the interplay of laser-induced photophoretic forces \cite{Du:2017,Wieben:2018} and plasma-induced asymmetric ion momentum transfer \cite{Krasheninnikov:2024}, whose relative contributions depend on experimental conditions \cite{Nosenko:2020PRR_JP,Nosenko:2022}. As a result, Janus particles exhibit persistent non-thermal motion and strongly affect the collective behavior of the suspension, for example by suppressing crystallization even when present in small concentrations \cite{Nosenko:25PRE}.

Complex plasmas with active Janus particles therefore combine several advantageous features for active-matter research. The strength of particle propulsion and the dissipation rate can be adjusted independently through the illumination intensity and gas pressure, allowing systematic control of activity and damping \cite{Nosenko:2020PRR_JP,Nosenko:2022}. At the same time, the low friction environment permits exploration of active dynamics in an inertial regime \cite{Loewen:2020,Martins:2022,Sprenger:2023,Lisin:2026}. This regime is difficult to access experimentally in most soft-matter realizations of active matter, which are typically overdamped. Another notable aspect of this system is that it enables the study of chaotic active flows in a stationary and approximately homogeneous setting. In contrast to classical inertial turbulence, which decays in the absence of sustained large-scale forcing \cite{Landau_v6}, activity continuously injects energy at the particle scale. Janus particles also provide a controlled realization of inhomogeneous particles in plasma, making them relevant for understanding the behavior of irregular or ``abnormal'' particles with complex trajectories \cite{Du:2017}.

In this paper, we investigate a two-dimensional (2D) complex plasma with active Janus particles, extending earlier studies to a higher level of activity. By increasing the illumination laser power, we access a highly energetic yet stable state that remains free of plasma-specific heating mechanisms such as the mode-coupling instability. Using particle-resolved diagnostics, we characterize single-particle motion and collective dynamics, and we analyze spatio-temporal correlations of the particle velocity field. The results demonstrate how complex plasmas with active Janus particles can serve as a model system for inertial active matter and provide insight into turbulence-like behavior in driven many-particle systems.

\section{Experimental method}

We used the experimental method of Ref.~\cite{Nosenko:25PRE}. The experiments were carried out in a modified Gaseous Electronics Conference (GEC) radio-frequency (rf) reference cell \cite{Hargis:1994,Couedel:2022}. Plasma was produced by a capacitively coupled $20$~W, $13.56$~MHz discharge in argon at the low pressure of $0.66$~Pa. The experimental conditions were as in Ref.~\cite{Nosenko:25PRE} except the illumination laser power, which was increased to $99$~mW. This resulted in higher mean kinetic energy of particles and more disordered state of their suspension compared with Ref.~\cite{Nosenko:25PRE}.

\begin{figure}[tbp!]
\centering
\includegraphics[width=1.0\columnwidth]{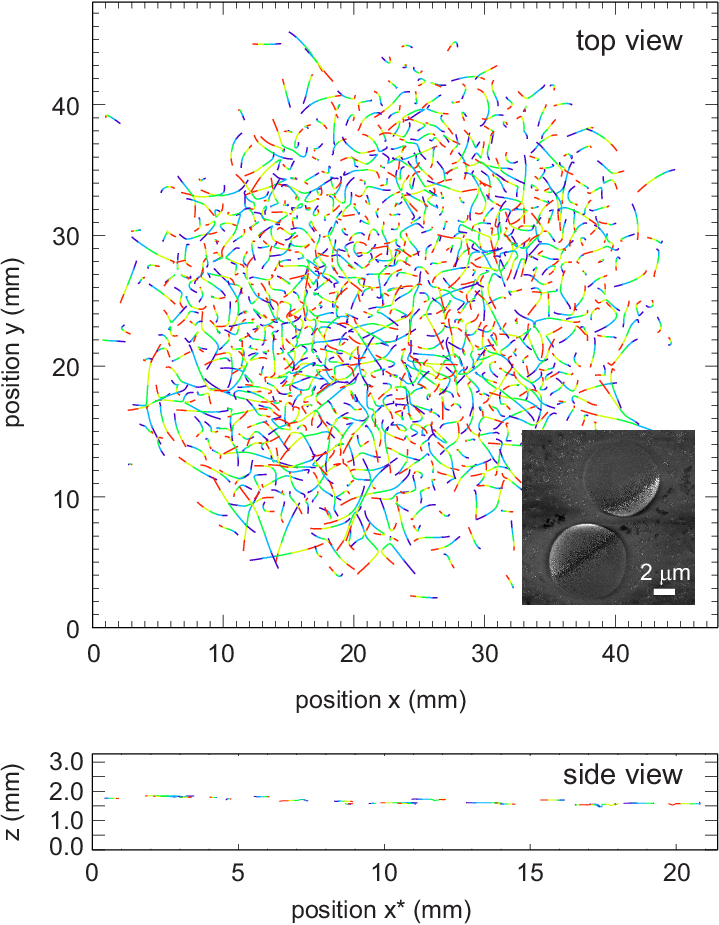}
\caption {\label {Fig_traj} Trajectories of Janus particles during $0.4$~s (time is color-coded from purple to red). (upper panel) Top view. The inset shows an SEM image of the Janus particles with $40$-nm caps of gold coating. (lower panel) Side view. Notice the different $x$ scales in the top- and side-view panels. The particles were suspended as a single layer in the plasma sheath above the lower powered electrode in a radio-frequency discharge in argon.
}
\end{figure}

To prepare the Janus particles used in this study, we started with melamine-formaldehyde (MF) microspheres \cite{microparticles} with a diameter of $9.27~\mu$m and mass $m_d=6.3\times10^{-13}$~kg. Using a drop-casting method described in Ref.~\cite{Nosenko:2020PRR_JP} and radio-frequency magnetron sputtering technique, they were coated on one side with a $\approx 40$~nm layer of gold. An SEM image of the resulting Janus particles is shown in the inset in the upper panel of Fig.~\ref{Fig_traj}.

After the particles were introduced into plasma from a manual dispenser, they were trapped at equilibrium position in the plasma sheath above the lower powered electrode. To the best of the author's knowledge, no direct measurements of the longevity of metal coating on plasma-suspended microparticles were reported in the literature. Indirect inferences were made in Ref.~\cite{Melzer:2025}, where a time scale of $60-80$~min was suggested from the observed trajectories of silver-coated MF microparticles suspended in argon plasma. Sputtering of the metal coating by impinging argon ions was suggested as the mechanism of coating degradation. It should be noted that the opposite process---redeposition of the material sputtered elsewhere in the discharge chamber onto the suspended particles---is also possible \cite{Kononov:2021}, which complicates the overall picture. To minimize potential damage to the particle surface from the plasma exposure, short waiting time on the scale of a few minutes (between the particle injection and video recording) was used in the present work.

A horizontal laser sheet with the wavelength of $660$~nm and a vertical laser sheet with the wavelength of $635$~nm were used to illuminate the particles. (The latter was used only for recording the side-view video and verifying that the particle suspension consisted of a single layer and was switched off during top-view video recording.) They were imaged from the top by the Photron FASTCAM Mini WX100 video camera and from the side by the Sony XC-ST50 camera. Bandpass interference filters matching the respective wavelengths of the illumination lasers were installed in the camera lenses to reject the undesired background light. The main experimental data was the top-view video of the particle motion. It was recorded at the rate of $125$ frames per second to the full capacity of the camera's onboard memory of $2726$ frames and the total duration of $21.8$~s. The camera's field of view was $47.84\times47.84~{\rm mm}^2$ ($2048\times2048~{\rm pixel}^2$), the pixel size was $23.36~\mu$m.

Data analysis proceeded in the following way. In every frame of the top-view camera video, the images of individual particles were identified and their positions were measured with subpixel resolution using an intensity moment method \cite{SPIT}. By tracing the particles from frame to frame, their displacements and velocities were determined. From the particle positions and velocities, various spatio-temporal correlation functions were calculated.

\section{Results and discussion}

When the Janus particles were injected in plasma, they levitated in the plasma sheath above the lower powered electrode where they formed a round single-layer suspension, see Fig.~\ref{Fig_traj}. The suspension consisted of about $770$ particles and had a diameter of approximately $40$~mm. In the vertical direction, the force of gravity acting on the particles was balanced by the sheath electric field. In the horizontal direction, the particle confinement was due to the ambipolar electric field naturally present in plasma. The sheath electric field has an effect also on the {\it orientation} of a suspended Janus particle. Whether a particle's axis of symmetry (connecting its coated and uncoated poles) would orient itself along the electric field is, however, not clear. As was predicted theoretically and shown experimentally, see Ref.~\cite{Annaratone:2011} and references therein, the orientation of a non-spherically-symmetric particle (rod-like in that case) suspended in the plasma sheath depends on the ratio of the particle's induced dipole and quadrupole moments, which ratio and the particle orientation can be tuned by varying the gas pressure and discharge power. These results serve as an illustration and cannot be directly extended to the case of a spherical Janus particle. The optical setup of the present work does not allow to resolve the particle orientation {\it in situ}.

As in the previous experiments \cite{Nosenko:2020PRR_JP,Nosenko:2022,Nosenko:25PRE}, the Janus particles became active when suspended in plasma, i.e., they acquired self-propulsion. The particles were driven by the photophoretic force from the illumination laser and asymmetric ion-drag force \cite{Nosenko:2020PRR_JP,Nosenko:2022}. The horizontal component of the self-propulsion force was balanced by the neutral gas drag force \cite{Epstein:1924,Liu:2003}. The single-layer suspension of Janus particles---unlike regular MF particles in similar experimental conditions---did not self-organize in a regular lattice (plasma crystal), which corroborates the results of the previous experiments \cite{Nosenko:2020PRR_JP,Nosenko:25PRE}. Instead, the particles moved around with high speed greatly exceeding the thermal speed, see the upper panel in Fig.~\ref{Fig_traj}. As was previously reported in Ref.~\cite{Nosenko:2022}, the addition of even a small fraction of Janus particles to regular MF particles also prevented their crystallization.

Comparing with the experiment of Ref.~\cite{Nosenko:25PRE}, in the present experiment much higher degree of activity and mean kinetic energy $\langle E_k \rangle\approx 117 \pm 36$~eV of the Janus particles was achieved due to the higher illumination laser power (the error was calculated as standard deviation for particle ensemble). The increase in $\langle E_k \rangle$ is $1.8$-fold compared with Ref.~\cite{Nosenko:25PRE} and $7$-fold compared with Ref.~\cite{Nosenko:2006}, where a laser-manipulation method was used to heat a single-layer suspension of regular (uncoated) MF particles. Note that in Ref.~\cite{Nosenko:2022}, the opposite (declining) dependence of $\langle E_k \rangle$ on the illumination laser power $P_{\rm laser}$ was observed. The opposite trends are most probably explained by different experimental conditions in the present work and Ref.~\cite{Nosenko:25PRE} on the one hand ($40$-nm gold coating on particles, $0.66$~Pa of argon), and Ref.~\cite{Nosenko:2022} on the other hand ($10$-nm platinum coating, $1.66$~Pa of argon). These differences presumably resulted in different balances of the photophoretic force and asymmetric ion drag force acting on a particle (which forces may be oppositely directed and therefore partially cancel out). In the literature, there are other examples of both types of $E_k(P_{\rm laser})$ dependencies for metal-coated particles suspended in plasmas. A rising dependence was reported in Ref.~\cite{Melzer:2025}, while a declining dependence was reported in Ref.~\cite{Vasiliev:2023}.

Importantly, the highly heated state of the particle suspension was achieved while avoiding the mode-coupling instability (MCI) \cite{Couedel:2010,Couedel:2011}. A key signature of MCI---mixed polarization of wave modes \cite{Couedel:2011}---was not present, since the particle trajectories remained in the horizontal plane, i.e., no detectable out-of-plane motion was recorded \cite{footnote1}, see the lower panel in Fig.~\ref{Fig_traj}. Therefore, this system is suitable for modeling generic phenomena in 2D active matter while ruling out plasma-specific effects.

We observed distinct spatio-temporal correlations in the highly heated 2D complex plasma with Janus particles, as described below.

\begin{figure}[tbp!]
\centering
\includegraphics[width=1.0\columnwidth]{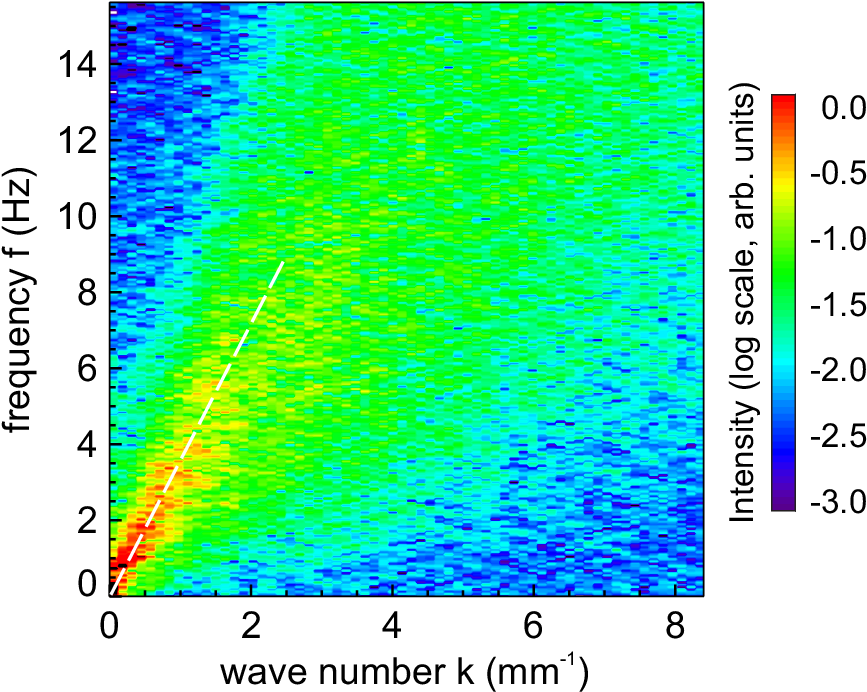}
\caption {\label {Fig_omega_k} Spectrum of the in-plane compressional waves in the single-layer suspension of Janus particles. Acoustic dispersion relation is evident ($\omega\propto k$ as indicated by the dashed line); the slope of it gives the speed of sound $C_s\approx 22.5$~mm/s.
}
\end{figure}

\begin{figure}[tbp!]
\centering
\includegraphics[width=0.9\columnwidth]{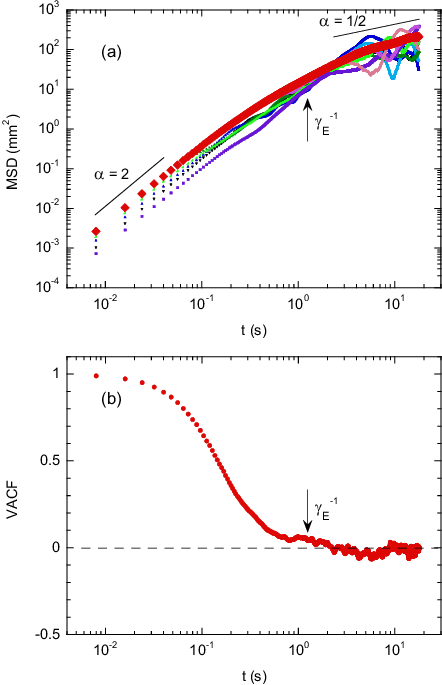}
\caption {\label {Fig_MSD_VACF} (a) Mean squared displacement MSD$(t)$ of $8$ selected particles (small symbols) and averaged over all particles (large diamonds). (b) Velocity autocorrelation function VACF$(t)$ averaged over all particles. The inertial delay time $\tau_m=\gamma_E^{-1}$ is shown by vertical arrows.
}
\end{figure}

{\it Waves.} Since the particle suspension was in a fluid state, it supported compressional in-plane wave mode. The wave spectrum of this mode was calculated using the method of Ref.~\cite{Nunomura:2002} (Fourier transform in space and time domains of the binned particle velocity field) and is shown in Fig.~\ref{Fig_omega_k}. The waves clearly have acoustic dispersion relation ($\omega\propto k$) without any detectable frequency- or wave-number cutoffs. On top of it, the spectrum is significantly broadened especially at higher wave numbers. The slope of the dispersion relation gives the speed of sound $C_s\approx 22.5$~mm/s. These features are compatible with the dust-thermal wave mode (DTW) in a liquid complex plasma \cite{Nunomura:2005}. The speed of the DTW mode is given by $v_{\rm DTW}=\sqrt{\gamma k_B T_d/m_d}$, where $\gamma$ and $T_d$ are respectively the adiabatic index and kinetic temperature of the dust component, $m_d$ is the dust particle mass, and $k_B$ is the Boltzmann constant \cite{Nunomura:2005}. Assuming $\gamma=2$ for the 2D particle suspension and $T_d=\langle E_k \rangle/k_B=1.36\times10^6$~K, this gives $v_{\rm DTW}\approx7.7$~mm/s, which is a factor of $2.9$ smaller than the measured sound speed $C_s\approx 22.5$~mm/s. A discrepancy this large probably points at the role of Janus particles' activity in sustaining these waves. At the same time, the transverse (shear) in-plane mode was not detected as expected for fluids \cite{Nunomura:2005}.

It is important to note that the in-plane compressional wave spectrum in Fig.~\ref{Fig_omega_k} does not feature MCI-specific hot spots \cite{Couedel:2011}. Together with the lack of out-of-plane particle oscillations in the lower panel in Fig.~\ref{Fig_traj}, this rules out MCI as a possible source of the particle suspension heating.

{\it Mean squared displacement}. To characterize the particle dynamics at the level of individual particles, we used their mean squared displacement defined as
\begin{equation}\label{MSD}
{\rm MSD}(t)=\langle|{\bf r}_i(t)-{\bf r}_i(t_0)|^2\rangle,
\end{equation}
where ${\bf r}_i(t)$ is the position of the $i$-th particle at time $t$. The brackets denote the average over $10$ different times $t_0$ separated by $0.4$~s and over all particles. ${\rm MSD}(t)$ is shown in Fig.~\ref{Fig_MSD_VACF}(a). It scales as ${\rm MSD}(t)\propto t^{\alpha}$ with $\alpha=2$, which is indicative of ballistic motion, at small times $t\ll\gamma_E^{-1}$, where $\gamma_E=0.8~{\rm s}^{-1}$ is the Epstein drag rate for particles \cite{Epstein:1924,Liu:2003,footnote2}. Here, the particle inertia is important \cite{Scholz:2018}. At later times, the dynamical exponent declines, briefly goes through the diffusive regime with $\alpha=1$ at $t\gtrsim\gamma_E^{-1}$, and finally reaches the value of $\alpha\approx0.5$.

Subdiffusive regime with $\alpha\approx0.5$ was also observed in Ref.~\cite{Nosenko:2022}. The underlying physics is not completely clear. Known situations where $\alpha=1/2$ include single-file diffusion \cite{Wei:2000}, which is obviously not the case here. Note that the ${\rm MSD}(t)$ functions of individual particles feature seemingly random oscillations especially for $t\gtrsim\gamma_E^{-1}$, see Fig.~\ref{Fig_MSD_VACF}(a). The dynamical exponent $\alpha\approx0.5$ is only revealed after averaging over all particles indicating the apparent emergence of collective particle dynamics driven by their propensity to move in circular trajectories, collisions, and external confinement.

{\it Velocity autocorrelation function}. This is another fundamental indicator calculated as
\begin{equation}\label{alpha}
{\rm VACF}(t)=\frac{\left\langle{\bf v}_i(t+t_0)\cdot {\bf v}_i(t_0)\right\rangle}
{\left\langle{\bf v}_i(t_0)\cdot {\bf v}_i(t_0)\right\rangle},
\end{equation}
where ${\bf v}_i(t)$ is the velocity of the $i$-th particle at time $t$. The brackets denote the average over $10$ different times $t_0$ separated by $0.4$~s and over all particles. VACF$(t)$ is shown in Fig.~\ref{Fig_MSD_VACF}(b). The initial exponential decay of it gives the interparticle collision time of $0.22$~s. VACF$(t)$ further declines up to the inertial delay time $\gamma_E^{-1}=1.25$~s and oscillates in the range of $\pm0.07$ at still greater times. This further illustrates how the collective particle dynamics emerges at $t\gtrsim\gamma_E^{-1}$, where the oscillations of individual particles cancel out.

\begin{figure}[tbp!]
\centering
\includegraphics[width=0.9\columnwidth]{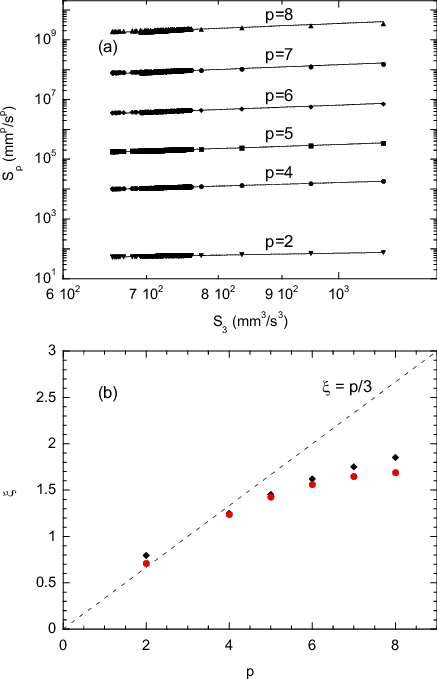}
\caption {\label {Fig_Sp} (a) Longitudinal velocity structure functions for a single-layer suspension of Janus particles plotted as $S_p(S_3)$. The power-law scaling of $S_p(S_3)$ reveals extended self-similarity of the particle velocity field. (b) Scaling exponent $\xi$ as a function of $p$ for the present experiment (red circles) and Ref.~\cite{Nosenko:25PRE} (black diamonds). The $\xi=p/3$ line indicates Kolmogorovian scaling. Adapted from V. Nosenko, Phys. Rev. E {\bf 111}, 045210 (2025); licensed under a Creative Commons Attribution 4.0 International license.
}
\end{figure}

{\it Extended self-similarity of the particle velocity field.} To quantify the emergence of collective dynamics beyond single-particle observables, we analyze the statistics of velocity fluctuations using longitudinal Eulerian structure functions. For a given spatial separation $r$, the structure function of order $p$ is defined as
\begin{equation}\label{Sn}
S_p(r)=\left\langle\left|({\bf v}_i-{\bf v}_j)\cdot {\bf r}_{ij}/r_{ij}\right|^p\right\rangle,
\end{equation}
where ${\bf r}_{ij}={\bf r}_i-{\bf r}_j$ and the averaging is performed over all particle pairs separated by $r$ and over time. In classical homogeneous turbulence, these quantities are predicted to scale as $S_p(r)\propto r^{\xi_p}$, with $\xi_p=p/3$ in the inertial range according to Kolmogorov theory \cite{Landau_v6}. In the present system, however, the functions $S_p(r)$ do not exhibit any clear power-law dependence on distance, reflecting the absence of a conventional inertial interval in real space.

Despite this, a robust scaling behavior emerges when structure functions of different orders are compared with each other. Specifically, plotting $S_p$ as a function of $S_3$ reveals extended self-similarity (ESS) \cite{Benzi:1993}, manifested by power-law relationships $S_p\propto S_3^{\xi_p}$ over a broad range of scales. The ESS representation of the structure functions for orders $p=2\!-\!8$ is shown in Fig.~\ref{Fig_Sp}(a). The existence of well-defined scaling in this representation indicates that, although spatial scaling is obscured by dissipation and activity at particle scales, the hierarchy of velocity fluctuations remains self-similar.

The scaling exponents extracted from the ESS plots are summarized in Fig.~\ref{Fig_Sp}(b). For lower orders ($p=2\!-\!4$), the exponents are close to the Kolmogorov prediction $\xi_p=p/3$, while at higher orders systematic deviations toward smaller values are observed. Such behavior is commonly associated with intermittency and multiscaling of the velocity field \cite{Nikora:2001}. Compared with the lower-activity experiment reported in Ref.~\cite{Nosenko:25PRE}, the deviations from $p/3$ are more pronounced here, consistent with the increased degree of activity and disorder. These results demonstrate that ESS persists in a highly driven, inertial active-matter system and that enhanced activity leads to stronger intermittency in the collective particle dynamics.

\begin{figure}[tbp!]
\centering
\includegraphics[width=0.9\columnwidth]{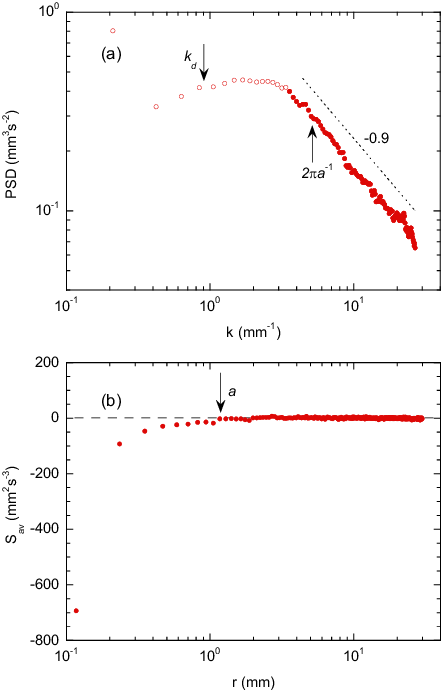}
\caption {\label {Fig_PSD} (a) Energy spectrum of the single-layer suspension of Janus particles. The solid symbols indicate the range of wave numbers where the energy cascade develops. (b) Crossed velocity-acceleration Eulerian structure function $S_{av}(r)$. The arrows indicate the mean interparticle spacing $a$, respective wave number $2\pi a^{-1}$, and the wave number $k_d$ corresponding to the inverse characteristic damping length.
}
\end{figure}

{\it Energy cascade}. The energy cascade is another hallmark of turbulence. In 3D high-Reynolds-number turbulence, energy is injected at the large (integral) spatial scales and flows down the inertial interval to smaller spatial scales, where it is dissipated by viscous friction \cite{Landau_v6}. This direct energy cascade is identified by the range of wave numbers $k$ where the energy spectrum $E(k)$ displays power-law scaling. Kolmogorov predicted the universal scaling exponent of $-5/3$ for the inertial interval of fully developed turbulence \cite{Landau_v6}. In 2D turbulence, the energy cascade is {\it inverse}, i.e., energy flows from smaller to larger spatial scales \cite{Kraichnan:1967}. In active turbulence, the energy cascade might take a different scaling or not develop altogether, since energy is injected and dissipated at similar spatial scales of individual particles \cite{Alert:2022}.

The energy spectrum of the Janus particle suspension was calculated as the Fourier transform of $S_2(r)$ and is shown in Fig.~\ref{Fig_PSD}(a). The energy cascade is evident in the range of wave numbers $k=3.6$--$26.7~{\rm mm}^{-1}$. As in Ref.~\cite{Nosenko:25PRE}, energy injection occurs at the spatial scales related to individual particles, as indicated by the position of the local maximum in $E(k)$ between the inverse damping length for particles $k_d=2\pi\gamma_E/\sqrt{k_B T_d/m_p}=0.9~{\rm mm}^{-1}$ and the inverse interparticle spacing $2\pi a^{-1}=5.2~{\rm mm}^{-1}$. Despite the 2D setting, the energy flow direction in the cascade is {\it direct}, as evidenced by the negative values, for $r<a$, of the crossed velocity-acceleration Eulerian structure function $S_{av}(r)=\langle\delta_r{\bf a}\cdot\delta_r{\bf v}\rangle$, where the brackets indicate averaging over all particle pairs and over time \cite{Bourgoin:2020}, see Fig.~\ref{Fig_PSD}(b).

Comparing the obtained energy spectrum with that in Ref.~\cite{Nosenko:25PRE}, one sees that the energy cascade in the present work follows the power-law dependence more closely. Here, the scaling exponent is approximately $-0.9$, which is larger than $-1.1$ reported in Ref.~\cite{Nosenko:25PRE}. Therefore, increased particle activity in the present experiment leads to a better-developed direct energy cascade with a larger scaling exponent.

Extended self-similarity of the velocity field and energy cascade are common signatures of turbulence. Finding out whether the chaotic active flow in a 2D system of Janus particles falls into the same universality class as known types of active turbulence is an open question and will require new dedicated experiments with varying degrees of activity.

To summarize, a 2D complex plasma with active Janus particles was studied experimentally. Increased illumination laser power resulted in enhanced Janus particle activity and greater degree of disorder in the system. However, it remained stable. In particular, the mode-coupling instability was not present as evidenced by the absence of detectable out-of-plane oscillations of particles and MCI-specific hot spots in the in-plane compressional wave spectrum. The measured speed of sound exceeded by a factor of $2.9$ the speed of the dust-thermal wave mode for given experimental conditions. This emphasizes the role of Janus particles' activity in sustaining these waves. The emergence of collective particle dynamics was shown using the particle mean squared displacement and velocity autocorrelation function. It was further shown that enhanced particle activity in this highly driven, inertial active-matter system leads to stronger intermittency in the collective particle dynamics and to a better-developed direct energy cascade while preserving extended self-similarity of the particle velocity field. These findings demonstrate that a complex plasma with active Janus particles is a promising model system to study collective phenomena in active matter in inertial regime.

\section{Acknowledgments}

Philip Born is acknowledged for producing the Janus particles used in this study. Matthias Kolbe is acknowledged for taking the SEM image of Janus particles. Till B\"{o}hmer is acknowledged for carefully reading the manuscript and helpful discussions.

\section{Author declarations}

The author has no conflicts of interest to disclose.

\section{Data availability}

The data that support the findings of this study are available from the corresponding author upon reasonable request.

\end{document}